\documentstyle[12pt,psfig,iopconf]{article}
\begin{document}
\title{Stellar Abundance Observations\footnote{To appear in the Proceedings
of the Second Oak Ridge Symposium on Atomic \& Nuclear Astrophysics}}

\author{John J Cowan\dag\footnote{cowan@mail.nhn.ou.edu}, 
Christopher Sneden\ddag\, James W Truran\S and Debra L Burris\dag
}

\affil{\dag\ 
Department of Physics and Astronomy, University of 
Oklahoma, Norman, OK 73019 }

\affil{\ddag\ 
Department of Astronomy, University of Texas, Austin, TX 78712
}

\affil{\S\ 
Department of Astronomy and Astrophysics, 
University of Chicago, Chicago, IL 60637}

\vspace{-14pt}
\beginabstract
Ground-  and space-based observations of stellar heavy element abundances
are providing a clearer picture of the  
chemical evolution of the Galaxy.
A large number of 
(r)apid and (s)low neutron capture process
elements, including the first Hubble
Space Telescope observations of 
Pt, Os, Pb and Ge, have been identified in 
metal-poor, galactic halo stars. 
In the very low metallicity (i.e. [Fe/H] $<$ --2.0) stars,
the abundance pattern of the elements from Ba 
through the third neutron-capture peak (Os-Pt) is 
consistent with a scaled solar {\it r}-process distribution. 
These results support previous observations that 
demonstrate the operation of the  {\it r}-process, including the synthesis
of the heaviest such elements, early in the history
of the Galaxy. 
New ground-based observations further confirm that  
the {\it s}-process element Ba   
and the {\it r}-process element 
Eu were both synthesized solely by the {\it r}-process at low metallicities,
and indicate the onset of the {\it s}-process occurred near [Fe/H] = --2.
Over a range of metallicity
 (--2.90 $<$ [Fe/H] $<$ --0.86) the data indicate that 
there exist real star-to-star differences in the ratios
 of the [n-capture/Fe] abundances as well as in the actual spectra of the
stars.

\endabstract

\vspace{-14pt}
\section{Introduction}
Observations of  elemental abundances in 
metal-poor halo stars provide important evidence regarding the 
early history, evolution and age of the Galaxy. 
Spectroscopic studies over a number of years have indicated 
the presence of 
neutron-capture, specifically rapid neutron-capture ({\it i.e.}
{\it r}-process), elements 
in 
a number of these metal-poor halo stars (see {\it e.g.} 
Spite and Spite 1978, Sneden and Parthasarathy 1983, Sneden 
and Pilachowski 1985,
Gilroy et al. 1988, Gratton and Sneden 1991, 1994, 
Sneden et al. 1994, McWilliam et al. 1995a, 1995b, 
Cowan et al. 1996,
Sneden et al. 1996, Burris et al. 1998).
In addition,
abundance comparisons of the {\it r}-process and {\it s}-process ({\it i.e.}
slow
neutron-capture) elements between the oldest metal-poor halo stars and 
more metal-rich halo and disk stars 
provide direct evidence about the nature of
the chemical evolution of the 
Galaxy.

\section{Abundance Observations of Metal-Poor Halo Stars}

The abundance patterns in the very old halo stars
indicate the nature of
the
early galactic populations and  nucleosynthetic processes. 
One of the most well-studied such stars is CS~22892--052. Although
iron poor 
([Fe/H]$ \simeq -$3.1\footnote
{We adopt the usual spectroscopic notations that
[A/B]~$\equiv$~log$_{\rm 10}$(N$_{\rm A}$/N$_{\rm B}$)$_{\rm star}$~--
log$_{\rm 10}$(N$_{\rm A}$/N$_{\rm B}$)$_{\odot}$, and that
log~$\epsilon$(A)~$\equiv$~log$_{\rm 10}$(N$_{\rm A}$/N$_{\rm H}$)~+~12.0,
for elements A and B. Also, metallicity will be arbitrarily defined
here to be equivalent to the stellar [Fe/H] value.}), 
it is neutron-capture element rich.
Sneden \etal\ (1996) have used high resolution, high 
signal-to-noise (S/N) spectra to examine the CS~22892-052 spectrum,
determining abundances of 20 neutron-capture elements in this star.
Some of these elements (such as terbium, holmium, thulium and 
hafnium) had never previously been detected in metal-poor halo stars.
We show in Figure~1 a comparison between the abundances of 
the neutron-capture elements in CS~22892--052 and a scaled 
solar {\it r}-process elemental
distribution (solid line). 

\vspace{-14pt}
\psfig{file=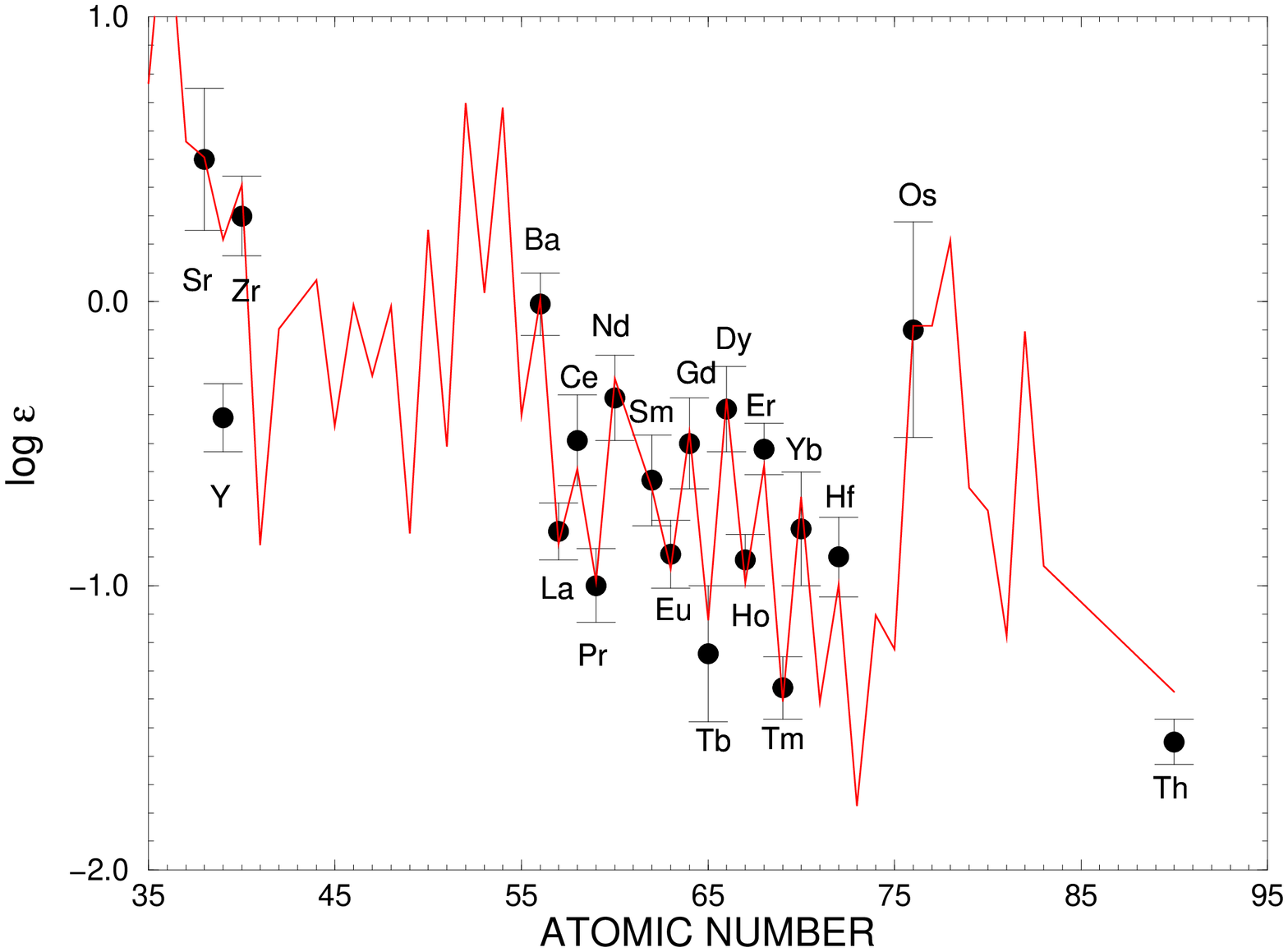,angle=-90,width=6truein}
\vspace{-14pt}
\par\noindent
Figure 1. Comparison of neutron-capture element abundances in 
CS 22892--052 with the solar system {\it r}-process
distribution.

While Os was observed in CS~22892--052, 
in general the 
3$^{rd}$ 
{\it r}-process peak elements (Os-Pt)
have dominant transitions in the uv, and thus are not accessible 
to ground-based observations. 
Recently we have observed three metal-poor halo stars using the GHRS of 
the Hubble Space Telescope (HST) (Sneden et al. 1998).
We show two of those stars, HD~115444 ([Fe/H] = --2.7) and HD~126238
([Fe/H] = --1.7), respectively in Figures 2 and  3.
The HST data are indicated by solid circles and 
ground-based data (an average of the observational values from
Griffin et al. 1982 and Gilroy et al. 1988) 
are indicated by open circles. For comparison we show 
in these figures a scaled total solar 
system (dashed line) and a solar system {\it r}-process (solid line)
abundance distribution. 

\psfig{file=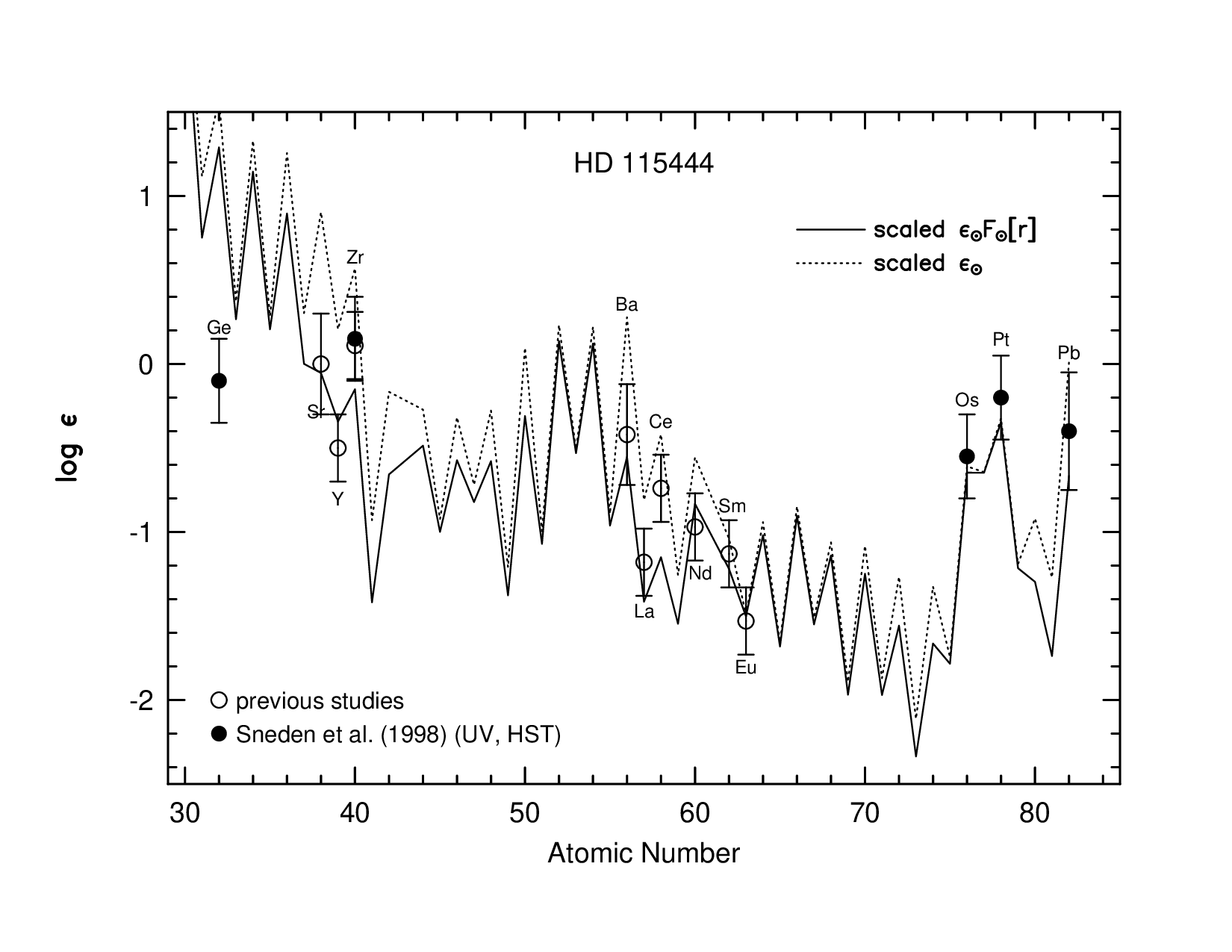,angle=0,width=5truein}
\vspace{-14pt}
\par\noindent
Figure 2. 
An abundance comparison between the neutron-capture elements in
HD~115444 ([Fe/H] = --2.7) 
and a scaled total solar system (dashed line) and solar system
$r$-process (solid line) abundance distribution.
Ground-based data (from various sources, see text for discussion) is
indicated by open circles, while HST data are indicated by solid circles.

\psfig{file=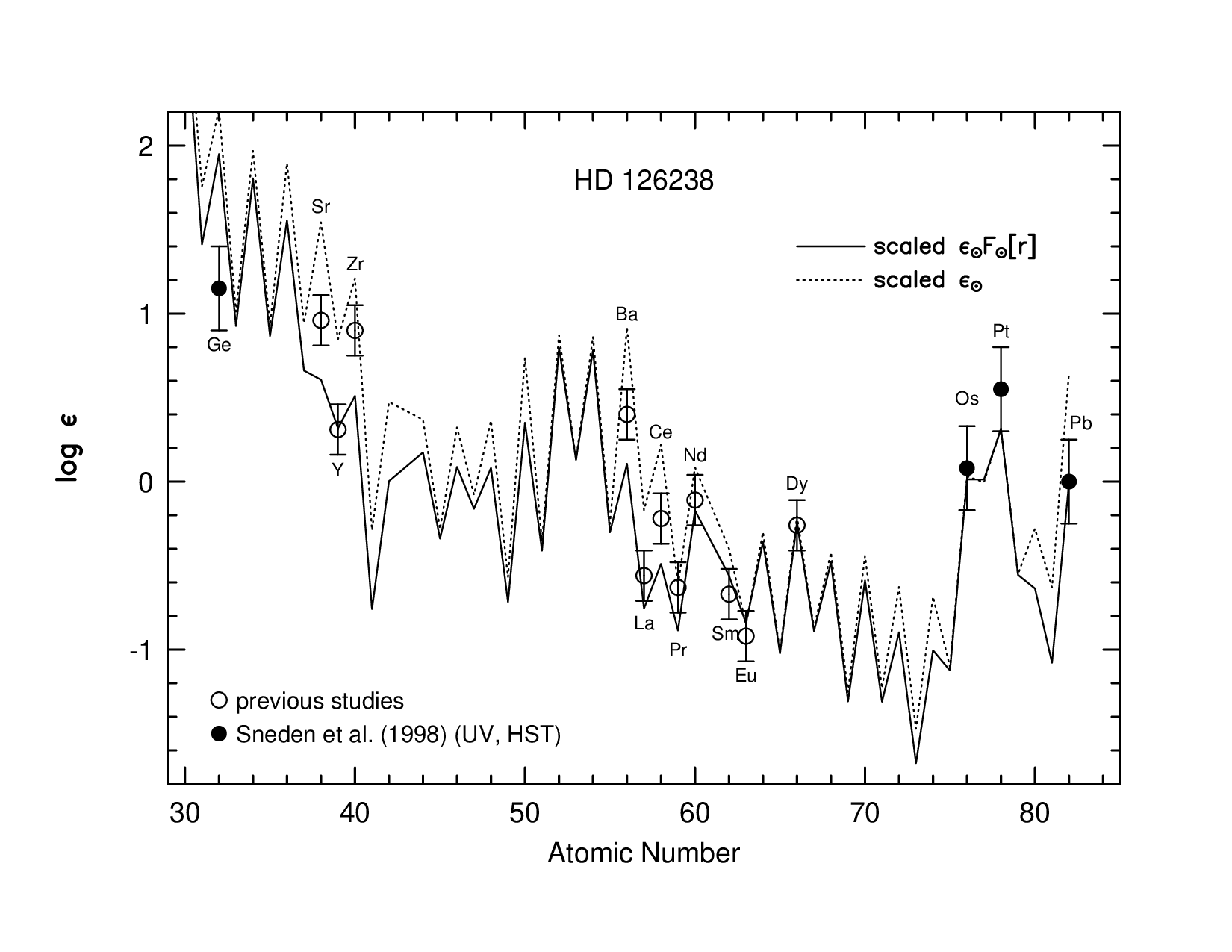,angle=0,width=5truein}
\vspace{-14pt}
\par\noindent
Figure 3. 
An abundance comparison for HD~122563 ([Fe/H] = --1.7) 
in the same style as that of Figure 2.

\section{Galactic Chemical Evolution Trends}

A study of a number of galactic halo stars, spanning a metallicty 
range of $-$2.96 to~$-$0.86, has recently been completed (Burris et al.
1998). Based upon data from Kitt Peak National Observatory 
neutron-capture element abundances have been obtained for approximately
40 stars. 
We show in Figure 4 the variation in [Ba/H] as a function of 
[$\alpha$/H], where $\alpha$ is the average of the Ca and Ti 
abundances. 
Included in the figure are the new stellar abundance 
data from Burris et al. (1998), 
indicated by solid circles.
Several other data sets for galactic halo stars are shown 
including those of Gratton and Sneden (1994) (open circles),
McWilliam et al. (1995a, b) (crosses) and Ryan et al. (1991, 1996) 
(triangles).
In addition, abundance studies of galactic disk stars by Edvardsson et al.
(1993) and Woolf et al. (1995) (periods) are shown in this
figure.
\vspace{-24pt}
\psfig{file=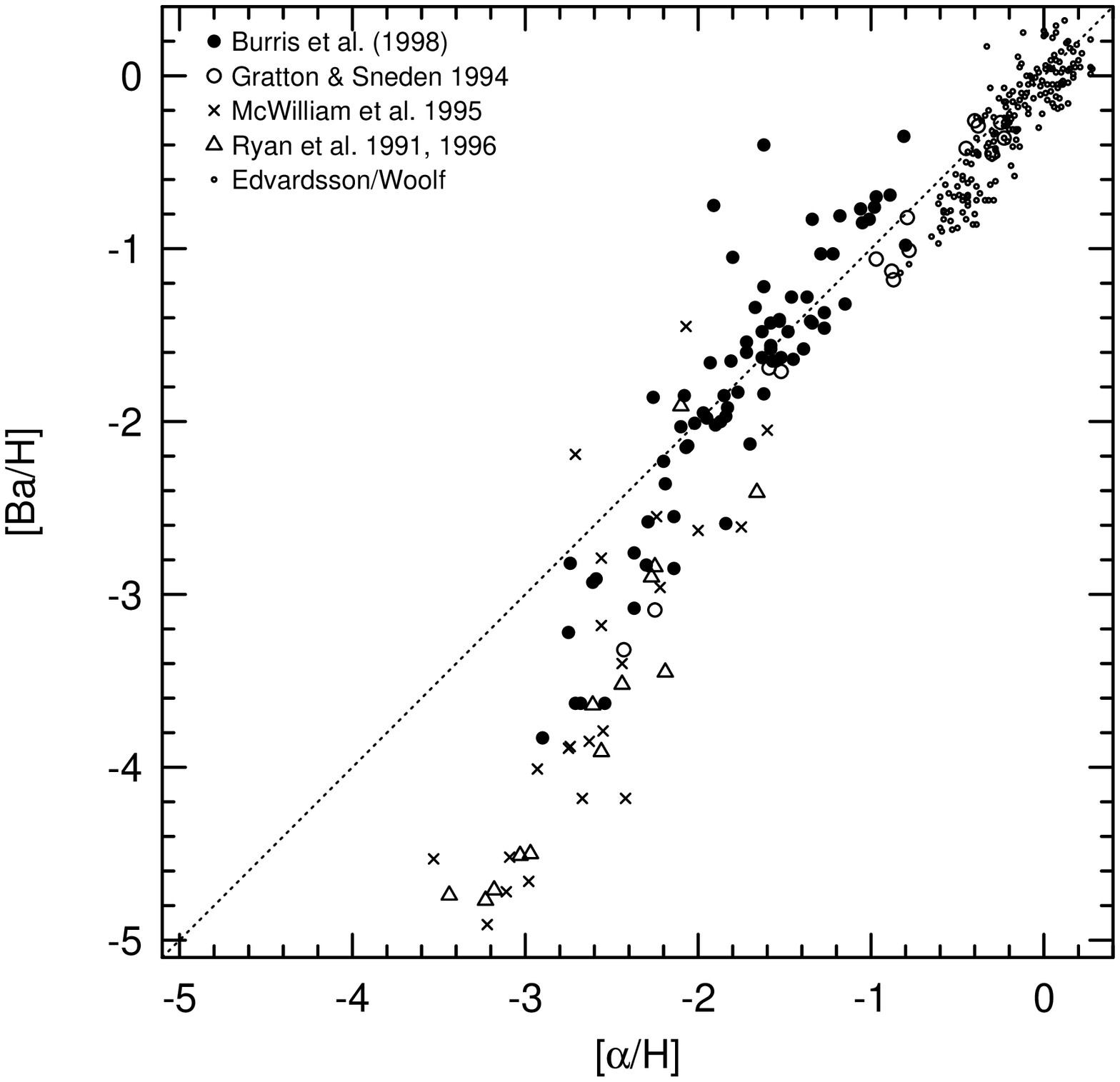,width=5truein}
\par\noindent
Figure 4. Comparison of [Ba/H] to [$\alpha$/H],
where $\alpha$ is the average of the Ca and Ti 
abundances, 
for a number of galactic stars.
\bigskip

As a comparison of the behavior of Ba, normally thought to be 
an {\it s}-process element, and Eu, normally thought to be an {\it r}-process
element, we plot the variation of [Ba/Eu] versus metallicity 
in Figure 5 for the same abundance sets as shown in Figure 4.

\psfig{file=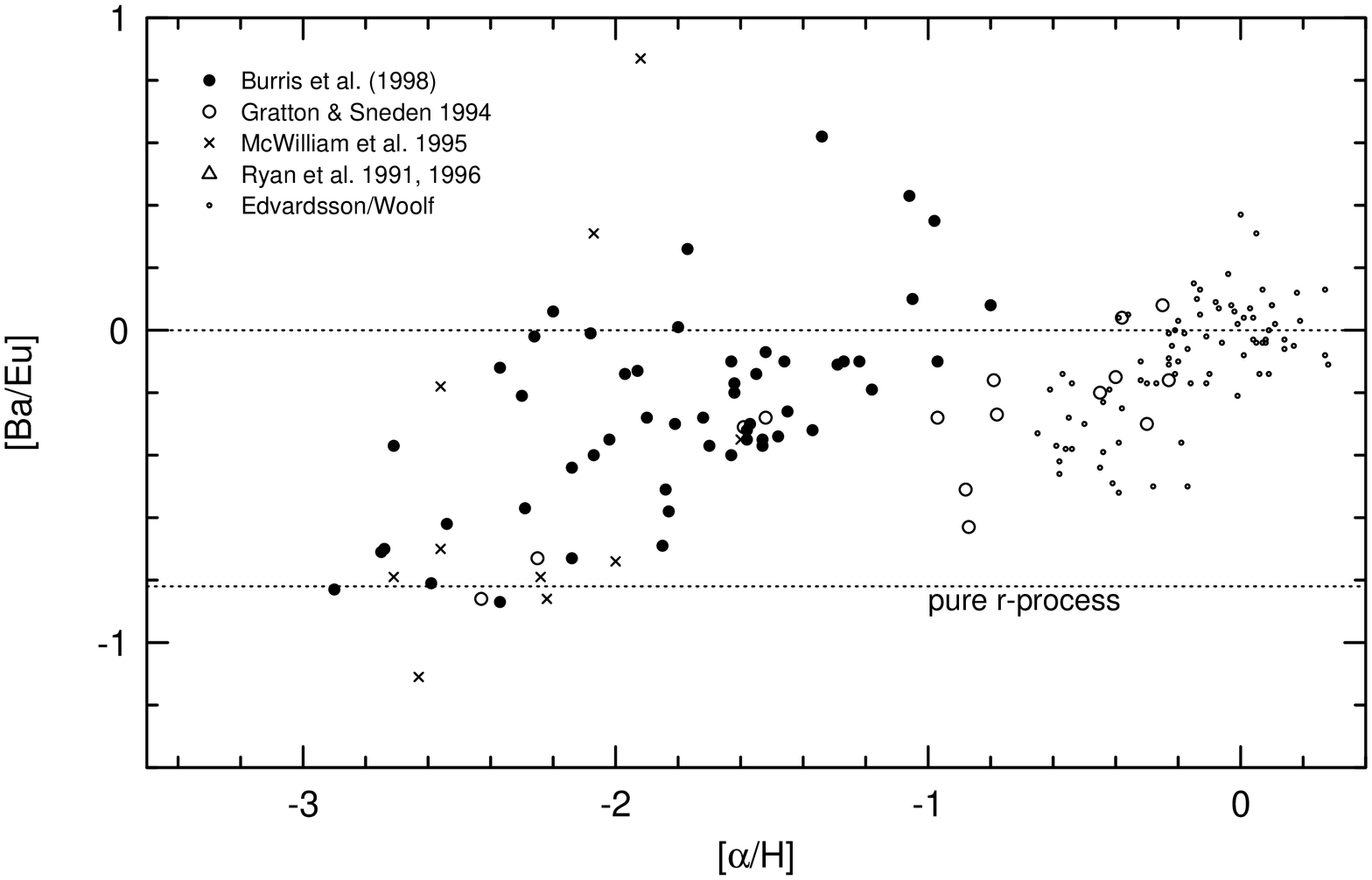,angle=90,width=5truein}
\vspace{-14pt}
\par\noindent
Figure 5. Comparison of [Ba/Eu] to [$\alpha$/H] for a number of galactic
stars.
\bigskip

Earlier work by Gilroy et al. (1988) 
presented evidence of a scatter 
in the abundances of the n-capture elements 
(with respect to iron) in the most metal-poor stars.
Using additional and more accurate data,
this suggestion has now been reexamined.
In Figure 6 we show the average neutron-capture element abundances, with
respect to iron, as a function of metallicity. 
For the data of Burris et al. (1998), the abundances of Ba, Eu, Nd, La 
and Dy  
were used. For all other data sets including McWilliam (1998),
the abundances of Ba, Nd  and Eu were used to determine the averages.

\psfig{file=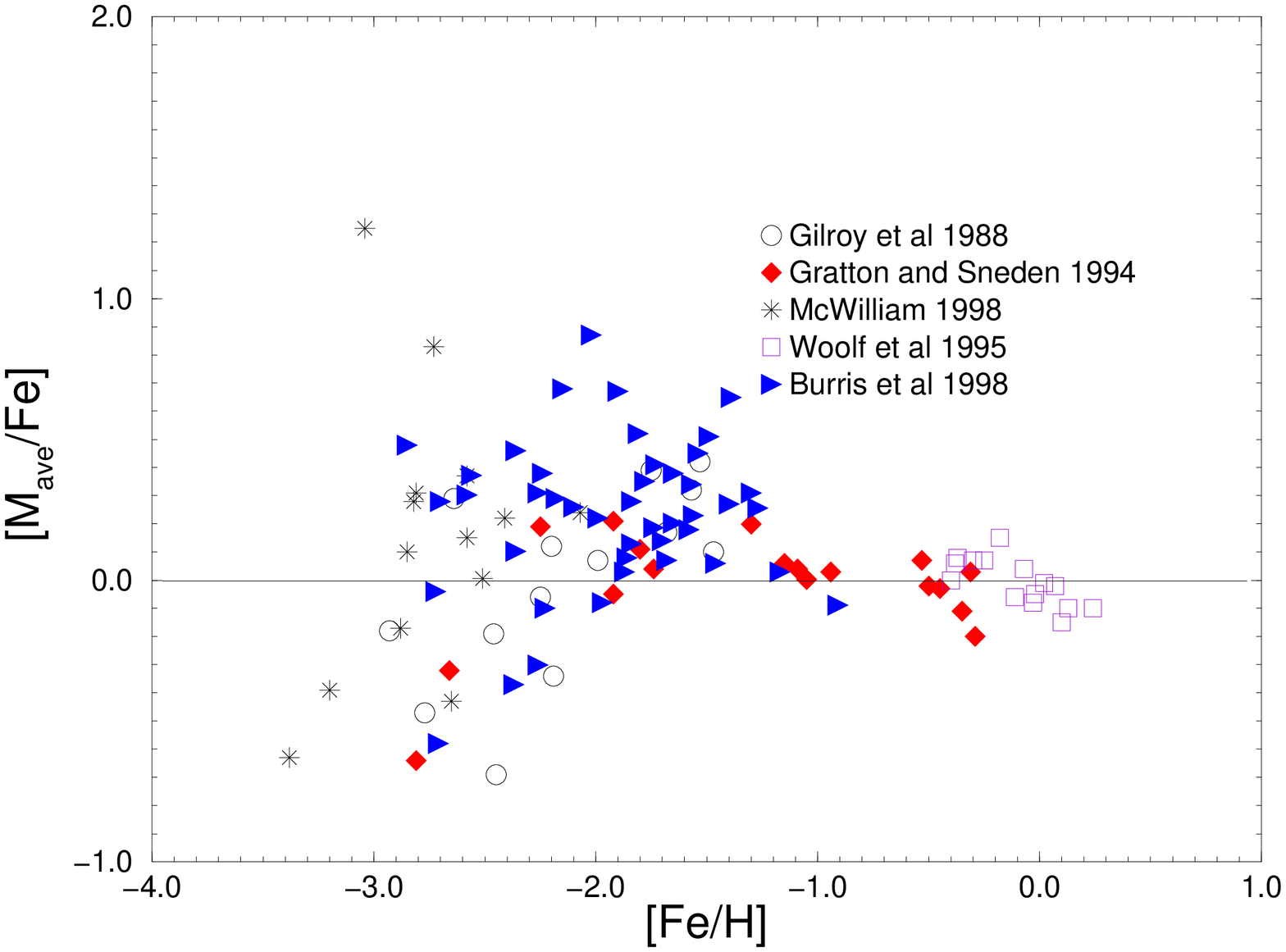,angle=-90,width=5truein}
\vspace{-14pt}
\par\noindent
Figure 6. Average abundances of n-capture elements/Fe as a function 
of metallicity.

\par
It is clear from Figure 6 that the newer data support the earlier
conclusions of Gilroy et al. (1988). At the lowest metallicities, there
is significant scatter in the total level of n-capture element/Fe
abundances. (The relative element-to-element abundances are the {\it same}
as shown above.) The  spectra of the stars clearly indicate 
large differences in the strength of the individual lines
(Burris et al. 1998).
It is seen in this figure that for increasing metallicity
the absolute scatter decreases and the average abundances 
for the most metal-rich halo stars and the disk stars are uniform.
The newer data also support the suggestion of Gilroy et al.
that at the earliest times ({\it i.e.} lowest metallicities) the Galaxy 
was not well mixed. Differences in the total n-capture element
abundances from star-to-star could then
be explained by the proximity of nucleosynthesis events (i.e. supernovae)  
prior to the formation of individual halo stars.

\bigskip

\section{Discussion and Conclusions}

Ground-based observations of the halo stars have indicated the presence of a 
number of neutron-capture elements. 
The availability of the
HST  has allowed for other spectral regions to be studied, and
as a result
more elements have been detected in these stars.
For example, Figures 2 and
3 illustrate (see also Sneden et al. 1998) 
that we have now detected the element Ge 
in (three) halo stars. 
In addition, the 3$^{rd}$ 
 {\it r}-process peak elements, Os-Pt, as well as Pb
have also
now been detected in two stars using the HST (Cowan et al. 1996,
Sneden et al. 1998).
Including the ground-based detections of Th in CS~22892--052
(see Figure 1)
and in HD 115444 (see Cowan et al. 1998), 
{\it r}-process
elements 
from  proton numbers of Z = 32 to 90 
have now been observed in metal-poor stars.
This is a much wider range in proton (and mass) number 
than ever seen before and now includes the important
3$^{rd}$ 
 {\it r}-process peak.
These detections further demonstrate that the {\it r}-process,
ranging up to the formation of the element Th, was in operation
early in the history of the Galaxy. 

The observations also provide important information about the 
nature of the progenitors of the halo stars. The {\it r}-process elements
cannot be 
internally synthesized in the halo stars. Instead they must be produced
in a previous generation (or generations) in an {\it r}-process site.
Since the metal-poor halo stars were formed early in the history
of the Galaxy, presumably shortly after formation,
the presence of {\it r}-process elements in the halo stars 
requires very short evolutionary
timescales for their progenitors. This further implies massive stars. 
While there is some uncertainty about the exact nature of 
the astrophysical site for the 
{\it r}-process, it has long been suspected to be in supernovae,
particularly core collapse supernovae from massive stars 
(see Cowan et al. 1991a). The abundance observations of the 
metal-poor halo stars appear to support that suspicion.

It has also become clear with the accumulating data 
that the neutron-capture elements 
in the metal-poor stars have an abundance pattern
that appears to be the same as the solar {\it r}-process distribution.
This is illustrated vividly in Figure 1, where all of the elements 
from Ba to Os in CS~22892--052 have relative solar abundances. 
While this correlation has been noted in the past
(see {\it e.g.} Gilroy et al. 1988),  the high-resolution
data in this star, covering a wide range of elements including
Os in the 3$^{rd}$ 
 {\it r}-process peak,
makes the argument
much stronger.
This same solar {\it r}-process pattern also appears in other stars, 
as shown in Figure 2.
Both ground-based and HST observations of HD~115444 ([Fe/H] = --2.7)
show that the  elemental abundance from Ba to Pt 
are  consistent with a scaled solar {\it r}-process curve.
Further evidence of this is seen in the metal-poor halo star
HD~122563 (Sneden et al. 1998). 
Elements such as Ba and La, which today are made predominantly in 
the {\it s}-process, appear to have been made exclusively in the {\it
r}-process
early in the history of the Galaxy. 
While this has been suggested previously (see {\it e.g.} Truran 1981),
the new observational data strongly support the contention that 
most (or all) of the elements were made in the {\it r}-process at the 
earliest times in the Galaxy. The observations indicate 
the same relative {\it r}-process abundance pattern in the oldest galactic
stars 
and in the solar material, at least for elements with 
Z $\ge$ 56.
Therefore, the data indicate that
the {\it solar system {\it r}-process abundances are not the result of 
global averages over different types of stars and epochs}. Instead, the 
stellar data
suggest that the conditions that produced the {\it r}-process 
elements are 
narrowly confined, perhaps both in terms of 
temperature and density of the nucleosynthesis 
and in terms of the mass range of the astrophysical sites
(see Wheeler et al. 1998, Freiburghaus et al. 1998). 
The apparent lack of mixing early in the history of the Galaxy,
when the relative abundance pattern is already apparent, 
demonstrated by Figure 6 also makes it less likely that the 
solar system {\it r}-process distribution is the result of 
averages.

We note in Figure 1, however, that while 
the elements in CS~22892--052 from Ba and above (Z $\ge$ 56) are well-fit
by the solar {\it r}-process distribution, the extrapolation to the lower
mass elements does not entirely fit the abundance data. 
In particular, while Sr and Zr do appear to be solar, the Y abundance
is far below the solar curve. It is difficult to explain why 
two but not three of these neighboring elements in this star are solar. 
We note, further, that the abundance data for HD~115444, shown in 
Figure 2, show a similar separation between the lighter and heavier
n-capture elemental abundances. In this star, again the abundances from 
Ba and above appear solar, but Sr-Zr do not. 
There may be several possible explanations. The weak {\it s}-process,
expected to occur during helium core burning in massive stars, 
is expected to contribute to the abundances of the elements from Sr-Zr.
The data may be showing such a contribution and Cowan et al.
(1995) even suggested some combination of the weak {\it s}-process and
the {\it r}-process might be needed to explain the abundances of Sr-Zr
in CS~22892--052. 
We note, however, one problem in this scenario is  the apparent
difficulty of producing {\it s}-process elements in stars of extremely
low metallicity.  
An alternative explanation may be that the more massive {\it r}-process
elements are synthesized in one site and the lower mass elements
in another. Based upon meteorite data, Wasserburg et al. (1996)
have suggested the existence of two {\it r}-process sites with the separation
in production occurring near mass number 140, {\it i.e.} near Ba. 
Possible alternative {\it r}-process sites have been discussed by
Wheeler et al. (1998) and Baron et al. (1998).
Further
observations and analyses will be needed to understand the formation
history of the lower mass {\it r}-process elements.

The most metal-rich of the halo stars studied here is 
HD~126238, with a metallicity of [Fe/H] = --1.7.
We see the same basic trends for this star in Figure 3
that are  seen in 
the more metal-poor stars CS~22892--052 and HD~115444. We note, however,
that the abundance of Ba seems to lie above the solar {\it r}-process curve.
It was suggested by Cowan et al. (1996) that this might indicate 
some {\it s}-process contribution to the Ba abundance. In other words,
by the time that HD~126238 formed at a metallicity of --1.7, 
presumably more recently 
than the other two previously mentioned stars, some galactic 
{\it s}-processing
had occurred. It is seen in Figure 3 that the stellar Ba abundance
is still below the total solar Ba abundance leading Cowan et al.
to suggest that only the most massive stellar contributors
to the {\it s}-process had evolved at that point in time.
Some support for their contention is given by Figure 4, which 
indicates the galactic chemical evolutionary trends for Ba as 
a function of metallicity. In this case metallicity is
indicated by $\alpha$,
which may be a more reliable metallicity indicator than Fe,
which is formed in both Type II and Type~I supernovae.
These data spanning a wide range in metallicty 
might be explained by an evolutionary delay in 
the production of {\it s}-process material. As demonstrated  earlier in
this paper,
at early times in the Galaxy Ba
apparently is produced from the {\it r}-process.
We note the clear change in slope in Figure 4 at 
a metallicity [$\alpha$/H] $\simeq$ --2
that appears to indicate the onset of the main {\it s}-process 
nucleosynthesis production for Ba 
(and presumably other n-capture elements)
in the Galaxy.
Further evidence of this change in production mechanism, as a function
of metallicity (and presumably time), for Ba is given in Figure 5.
At very low metallicities (and early times) the {\it s}-process 
element Ba and the {\it r}-process element Eu appear to be synthesized solely
in the {\it r}-process. While there is scatter in the available data,
we see that at the lowest values of [$\alpha$/H] the [Ba/Eu] value
in most of the stars is consistent with a pure {\it r}-process origin.   

The long-lived radioactive nuclei (known as chronometers)
in the uranium-thorium region
are formed exclusively by the {\it r}-process and can be used 
to determine the ages of stars and the Galaxy.
One such chronometer, Th, has been detected 
in CS~22892--052 (Sneden et al. 1996, Cowan et al. 1997) (see Figure 1).
Comparison between the initial abundance value produced in an {\it r}-process
site
(often known as the production value) 
and the observed abundance value leads to a direct estimate of
stellar ages. 
The abundances of the stable elements in the 3$^{rd}$ {\it r}-process peak,
a nuclear region nearby to the U-Th region,
can be used to help constrain the predictions of the  initial
values of the long-lived radioactive chronometers,
independent of knowing the site for the {\it r}-process.

Comparing the solar  and the observed  Th/Eu ratio
in CS 22892--052, Cowan et al. (1997) found an age 
estimate of 15 $\pm$ 4 Gyr for this star. 
They noted that 
consideration of galactic chemical evolution could lead to 
an older age 
of 17 $\pm$ 4 Gyr. 
Pfeiffer et al. (1997) employed  newer and more accurate nuclear data
in the context of a waiting point approximation {\it r}-process model.
Using the stable stellar and solar data to constrain
the predicted abundances of the radioactive {\it r}-process nuclei,
they compared the initial (as opposed to solar) value of Th/Eu 
with the stellar value and found a best estimate for
this star of 13.5 Gyr. Their result for CS~22892--052 
was not only consistent with
Cowan et al. (1997), but 
is also consistent with recent globular cluster
age determinations based upon Hipparcos data (see Pont et al. 1998).
(See also Cowan et al. 1991a,b for a discussion of galactic 
and cosmological age determinations.)

This technique,
based upon predicted and observed radioactive
chronometers, 
has been extended to an additional
star with an age result approximately the same as
that for CS~22892--052 (Cowan et al. 1998).
We caution, however, that there are still many uncertainties,
and to improve the accuracy of the chronometric estimates
will require more observational and theoretical
studies.
It is encouraging to note, though, that the 
detection of thorium in CS 22892--052, and other 
halo stars, offers promise 
as an independent technique
for determining stellar ages, and thus putting limits on galactic
and cosmological age estimates.

\section*{Acknowledgments}

Support for this work was provided through grants GO-05421, GO-05856, 
and GO-06748 from the Space Science Telescope Institute, which is 
operated by the Association of Universities for Research in Astronomy,
Inc., under NASA contract NAS5-26555. 
Support was also provided by the National Science Foundation 
(AST-9618332 to J.J.C.)    
and (AST-9618364 to C.S.).


\end{document}